\documentclass[preprint,10pt]{elsarticle}
\usepackage{amssymb}
\usepackage{amsmath}
\usepackage{graphicx}
\usepackage{lineno}
\biboptions{comma,square}

\biboptions{super}

\begin{document}

\begin{frontmatter}

\title{Scaling Properties of light (anti)nuclei and (anti)hypertriton
production\\ in Au+Au collisions at $\sqrt{s_{\rm{NN}}} = 200$~GeV}

\cortext[$^*$]{corresponding author}
\author{Chen Gang\corref{$^*$}}\ead{chengang1@cug.edu.cn}
\author{Chen Huan, Wang Jiang-Ling and Chen Zheng-Yu}
\address{
 School of Mathematics and Physics, China University of Geosciences, Wuhan
430074, China.}

\begin{abstract}
We present the scaling properties of mass number of light
(anti)nuclei production in midrapidity Au + Au collisions at $\sqrt
{s_{NN}}=200$~GeV based on the {\footnotesize PACIAE + DCPC} model.
It is found that the integrated yield of light (anti)nuclei
decreased exponentially with the increase of mass numbers which
depends on the centrality, this properties of the system can be
described quantitatively by temperature $T$ at hadronic freeze-out,
and the model results are consistent with STAR data. Furthermore, we
found that the integrated yield of heavier (anti)nuclei per
participant nucleon increases from peripheral to central collisions
more rapidly than that of $d(\bar{d})$, indicating that the mass
scale of light (anti)nuclei production was presented in relativistic
heavy ion collisions. .
\end{abstract}

\end{frontmatter}


\section{Introduction}

The hot and dense matter formed in relativistic heavy ion collision
experiments, which is similar to the fireball environment  in the
initial stages of the big bang, provides a good opportunity for
studying the production of light
(anti)nuclei.~\cite{prl2005,star2007}. The STAR Collaboration has
reported their measurements of $^3_{\Lambda}H$,
$\overline{_{\overline\Lambda}^3H}$ and $\overline{^4He}$ in Au+Au
collisions at the top energy available at the BNL Relativistic Heavy
Ion Collider (RHIC)~\cite{star2,star3}. The ALICE Collaboration has
also published its preliminary $\overline d$ yield of $\sim6\times
10^{-5}$ measured in the proton-proton collisions at $\sqrt s=7$~TeV
~\cite{alice,alice1}.

In theory, light (anti)nuclei can be produced via two mechanisms.
The first mechanism is direct production of nucleus-antinucleus
pairs in elementary nucleon-nucleon or parton-parton interactions.
However, because of their small binding energies, such (anti)nuclei
are likely to be dissociated in the medium before escaping. The
second, and presumably dominant, mechanism for antinucleus
production is via final-state coalescence~\cite{jili1,jili2,jili3}.
The light (anti)nuclei and (anti)hypernuclei are usually studied by
the transport models + the phase space coalescence
model~\cite{grei,chen,ma} and/or the statistical
model~\cite{pop,peter,ma1,stei} etc. For example, their production
rate can be described by both thermodynamic model and coalescence
model. In thermodynamic model, the system created is characterized
by the chemical freeze-out temperature $(T_{ch})$, kinetic
freeze-out temperature $(T_{kin})$, as well as the baryon and
strangeness chemical potential $\mu_B$ and $\mu_S$. An (anti)nucleus
is regarded as an object with energy $E_A = Am_p (A$ is the atomic
mass number, $m_p$ is the mass of proton) emitted by the fireball
~\cite{peter}. The production rates are proportional to the
Boltzmann factor $e^{-m_pA/T}$~\cite{star3}. In a microscopic
picture, a light (anti)nucleus is produced during the last stage of
the collision process via the strong correlation between the
constituent nucleons in their phase
space~\cite{0909,ma51,ma52,prc2002}.

Recently, we have proposed an approach the dynamically constrained
phase-space coalescence model ({\footnotesize DCPC})~\cite{yuyl}
which is based on the final hadronic state generated by a parton and
hadron cascade model PACIAE~\cite{sa2}. Using this method, the light
nuclei (anti-nuclei) yields, transverse momentum distribution, and
the rapidity distribution in non-single diffractive proton-proton
collisions at $\sqrt{s }=7$~TeV~\cite{yuyl} are predicted, and the
light nuclei (anti-nuclei) and hypernuclei (anti-hypernuclei)
productions~\cite{Cheng} and its centrality dependence~\cite{Cheng2}
in the Au+Au collisions at $\sqrt{s_{\rm{NN}}}=200$~GeV are
investigated.

Previous studies have shown that light nuclei production depends on
the mass number $A$ in Ultrarelativistic Heavy-Ion
Collisions~\cite{prl83}, and yields of light (anti)nuclei and
(anti)hypertriton per participant nucleon (Npart) depends on the
mass number in Au+Au collisions at $\sqrt{s_{\rm{NN}}}=200$
GeV~\cite{Cheng}. In this paper, we study the yields of light
(anti)nuclei and (anti)hypertriton depending on the mass number in
Au+Au collisions at $\sqrt s_{NN} = 200$~GeV with the {\footnotesize
PACIAE + DCPC} model in an attempt to shed light upon the possible
production mechanisms.

\section {Models}

The parton and hadron cascade model {\footnotesize
PACIAE}~\cite{sa2} is based on {\footnotesize PYTHIA}
6.4~\cite{sjo2} and is devised mainly for the nucleus-nucleus
collisions. In the {\footnotesize PACIAE} model, firstly, the
nucleus-nucleus collision is decomposed into the nucleon-nucleon
($NN$) collisions according to the collision geometry and $NN$ total
cross section. Each $NN$ collision is described by the
{\footnotesize PYTHIA} model with the string fragmentation switching
off and the diquarks (antidiquarks) randomly breaking into quarks
(anti-quarks). So the consequence of a $NN$ collision is a partonic
initial state composed of quarks, anti-quarks, and gluons. Provided
all $NN$ collisions are exhausted, one obtains a partonic initial
state for a nucleus-nucleus collision. This partonic initial state
is regarded as the quark-gluon matter (QGM) formed in relativistic
nucleus-nucleus collisions. Second, the parton rescattering
proceeds. The rescattering among partons in QGM is randomly
considered by the 2$\rightarrow$ 2 LO-pQCD parton-parton cross
sections~\cite{comb}. In addition, a $K$ factor is introduced here
to account for higher order and non-perturbative corrections. Third,
hadronization happens after parton rescattering. The partonic matter
can be hadronized by the Lund string fragmentation
regime~\cite{sjo2} and/or the phenomenological coalescence
model~\cite{sa2}. Finally, the hadronic matter continues
rescattering until the hadronic freeze-out (the exhaustion of the
hadron-hadron collision pairs). We refer to~\cite{sa2} for the
details.

In quantum statistical mechanics~\cite{kubo} one can not precisely
define both position $\vec q\equiv (x,y,z)$ and momentum $\vec
p\equiv (p_x,p_y,p_z)$ of a particle in the six-dimension phase
space, because of the uncertainty principle $\Delta\vec q\Delta\vec
p \geqslant h^3.$ We can only say that this particle lies somewhere
within a six-dimension quantum ¡°box¡± or ¡°state¡° with a volume of
$\Delta\vec q\Delta\vec p$. A particle state occupies a volume of
$h^3$ in the six-dimension phase space~\cite{kubo}. Therefore one
can estimate the yield of a single particle by defining an integral
$Y_1=\int_{H\leqslant E} \frac{d\vec qd\vec p}{h^3}$, where $H$ and
$E$ are the Hamiltonian and energy of the particle, respectively.
Similarly, the yield of N particle cluster can be estimated as
following integral~\cite{yuyl,Cheng}
\begin{equation}
Y_N=\int ...\int_{H\leqslant E} \frac{d\vec q_1d\vec p_1...d\vec
q_Nd\vec p_N}{h^{3N}}. \label{phas}
\end{equation}
In addition, the Eq.~(1) must satisfy the constraint condition
following
\begin{equation}
     m_0\leqslant m_{inv}\leqslant m_0+\Delta m,\\
\end{equation}
\begin{equation}
     q_{ij}\leqslant D_0, \ \ (i\neq j;\ i,j=1,2,\cdots,N).
\end{equation}
Where,
\begin{equation}
 m_{inv}=\Bigg[\bigg(\sum_{i=1}^N E_i\bigg)^2-\bigg(\sum_{i=1}^N
\vec p_i\bigg)^2\Bigg]^{1/2}, \label{yield2}
\end{equation}
and $E_i,\vec p_i (i=1,2,\cdots,N)$
are the energies and momenta of particles, respectively. $m_0$ and
$D_0$ stands for, respectively, the rest mass and diameter of light
(anti)nuclei, $\Delta m$ refers to the allowed mass uncertainty, and
$ q_{ij}=|\vec q_{i}-\vec q_{j}|$ is the vector distance between
particle $i$ and $j$. Because the hadron position and momentum
distributions from transport model simulations are discrete, the
integral over continuous distributions in Eq.~(1) should be replaced
by the sum over discrete distributions~\cite{Cheng2}.

\section{The mass number dependence of light (anti)nuclei production at RHIC}

 First we produce the final state particles using the
{\footnotesize
 PACIAE} model. In the {\footnotesize PACIAE} simulations we assume that hyperons are
heavier than $\Lambda$ decayed already. The model parameters are
fixed on the default values given in {\footnotesize
PYTHIA}~\cite{sjo2}. However, the $K$ factor as well as the
parameters parj(1), parj(2), and parj(3), relevant to the strange
production in {\footnotesize PYTHIA}~\cite{sjo2}, are given by
fitting the STAR data of $\Lambda$, $\overline \Lambda$, $\Xi^-$,
and $\overline{\Xi^-}$ in Au+Au collisions at $\sqrt{s_
{\rm{NN}}}=200$~GeV~\cite{star5}. The fitted parameters of $K$=3
(default value is 1 or 1.5~\cite{sjo2}), parj(1) = 0.12 (0.1),
parj(2) = 0.55 (0.3), and parj(3) = 0.65 (0.4) are used to generate
$1.02\times 10^8$  minimum-bias events by the {\footnotesize PACIAE}
model for Au+Au collisions at $\sqrt{s_{\rm{NN}}}=200$~GeV with
$|y|<1$ and $0 < p_t<5$ acceptances~\cite{Cheng}.

Then, the yields $dN/dy$ of $d$ ($\overline d$), $^3{He}$
($^3{\overline{He}}$), $_{\overline\Lambda}^3H$
($\overline{_{\overline\Lambda}^3H}$), as well as  $^4{He}$
($^4{\overline{He}}$) are calculated by the {\footnotesize DCPC}
model for different centrality bins of 0-5\%, 5-10\%, 10-15\%,
15-20\%, 20-30\%, 30-40\%, and 40-60\%, as shown in
Tab.~\ref{paci1}. We can see from Tab.~\ref{paci1} that the yields
$dN/dy$ of light (anti)nuclei and (anti)hypertritons decrease (or
increase) with the increase of centrality (or $N_{\rm {part}}$); the
yields of antinuclei are less than those of their corresponding
nuclei; and the greater the mass is, the lower the yield is.

\begin{figure}[htbp]\begin{center}
\includegraphics[width=0.65\textwidth]{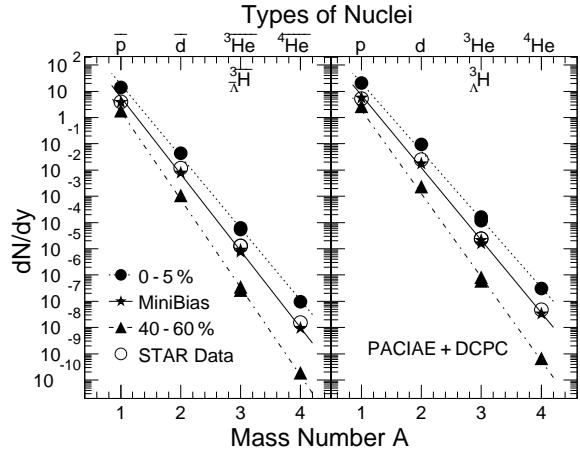}
\caption{Atomic number dependence of the integrated yield $dN/dy$ of
light (anti)nuclei ($d, \overline d$, $_{\Lambda}^3H,
\overline{_{\overline\Lambda}^3 H}$, $^3{He}, ^3{\overline{He}}$,
$^4{He}$, $^4{\overline{He}}$) in different centrality bins of
0-5\%,40-60\%, and MiniBias, respectively. The solid symbols are
calculated by {\footnotesize PACIAE+DCPC} model in the midrapidity
Au+Au collisions at the RHIC energy, while the open ones represent
the data points extracted by the STAR experiment~\cite{0909,star3}.
The lines represent the Model result's fit for the positive matters
(right) and antimatters (left) with formula $e^{-Am_p/T}$.}
\label{tu1}\end{center}
\end{figure}

In order to further explore the mass and/or centrality dependence of
light (anti)nuclei production, we study the integrated yield $dN/dy$
of light (anti)nuclei as a function of atomic mass number $A$ and/or
centrality. Fig.~\ref{tu1} shows the distributions of integrated
yield $dN/dy$ of light (anti)nuclei ($d, \overline d$,
$_{\Lambda}^3H, \overline{_{\overline\Lambda}^3 H}$, $^3{He},
^3{\overline{He}}$, $^4{He}$, $^4{\overline{He}}$) vs atomic mass
number $A$ with $A =1$ to $A =4$ in different centrality bins of
0-5\%, 40-60\%, and MiniBias, respectively.
\begin{table*}[t]
\begin{center}
\caption{Integrated yields $dN/dy$ calculated by {\footnotesize
PACIAE+DCPC} model of $\overline d$, $d$,
$\overline{_{\overline\Lambda}^3 H}$, ${_{\Lambda}^3 H}$,
${\overline{^3 He}}$, $^3{{He}}$, ${\overline{^4 He}}$ and
$^4{{He}}$ in midrapidity Au+Au collisions of
$\sqrt{s_{\rm{NN}}}=200$~GeV. $\langle N_{part}\rangle$ is shown for
each centrality.} {\footnotesize
\begin{tabular}{ccccccccc}
\hline  \hline \cline{1-8} Centrality
&$0-5\%$&$5-10\%$&$10-15\%$&$15-20\%$&$20-30\%$&$30-40\%$ &
$40-60\%$ \\ \hline
$\langle N_{part}\rangle$&334&  295&  252&  213&  164&  108&  54\\
$d^a$&0.0964 & 0.0837  & 0.0537 &  0.0389 & 0.0224 &  0.00936&   0.00239 \\
$\overline{d}^a$&0.0433 & 0.0383&  0.0244 & 0.0179 & 0.0101&   0.00425&  0.00107\\
${^3He}^b$ &1.60E-04&  1.17E-04 & 6.52E-05&  3.76E-05 &  1.60E-05&  3.78E-06&  8.19E-07\\
${\overline{^3 He}}^b$ &6.52E-05 & 4.88E-05 & 3.05E-05 & 1.69E-05 & 7.23E-06 & 1.71E-06  &3.53E-07 \\
$_{\Lambda}^3 H^b$ &1.21E-04&9.11E-05&  4.95E-05 & 2.97E-05 & 1.28E-05 &  2.90E-06 & 6.07E-07 \\
${\overline{_{\overline\Lambda}^3 H}}^b$&5.59E-05 & 4.33E-05 & 2.43E-05 & 1.43E-05 & 6.33E-06 & 1.39E-06 & 2.71E-07  \\
${^4He}^c$ &5.14E-07&  3.22E-07 & 1.44E-07&  9.36E-08 &  2.27E-08&  3.90E-09&  1.13E-09\\
${\overline{^4 He}}^c$ &1.70E-07 & 7.45E-08 & 4.22E-08 & 2.46E-08 & 7.59E-09 & 1.16E-09  &3.32E-10\\
\hline \hline \multicolumn{8}{l}{$^a$ calculated with $\Delta
m=0.00035$~GeV for $d,\overline{d}$.} \\ \multicolumn{8}{l}{$^b$ calculated with $\Delta m=0.00055$~GeV for $^3He,{\overline{^3 He}},_{\Lambda}^3 H,{\overline{_{\overline\Lambda}^3 H}}$.} \\
\multicolumn{8}{l}{$^c$ calculated with $\Delta m=0.0007$~GeV for $^4He,{\overline{^4 He}}$.} \\
\\\end{tabular}} \label{paci1}
\end{center}
\end{table*}
The solid symbols are calculated by {\footnotesize PACIAE+DCPC}
model in the midrapidity Au+Au collisions at the RHIC energy, while
the open ones represent the data points extracted from the STAR
experiment~\cite{0909,star3}. Obviously, the model results are
consistent with STAR data~\cite{0909,star3}, indicating that the
selection parameter $\Delta m$ calculated yields of light
(anti)nuclei in Tab.~\ref{paci1} are appropriate.

One can see from Fig.~\ref{tu1}, that the yields of light
(anti)nuclei and (anti)hypertriton all decrease rapidly with the
increase of atomic mass number $A$, and the integrated yield $dN/dy$
span almost 5 orders of magnitude with striking exponential
behavior. In this figure the curve is fitted to the data point using
equation as~\cite{prl83,plb1995,prl1998nu}
\begin{equation}
E_A\frac{d^3N_A}{d^3P_A}\varpropto E_Ae^{-m_pA/T},
\end{equation}
where $E_A\frac{d^3N_A}{d^3P_A}$ stands for the invariant yield of
(anti)nuclei, $P_A$ is the momentum of (anti)nuclei, and $T$ is the
temperature at hadronic freeze-out. The temperature parameters $T$,
fitting to the integrated yield $dN/dy$ of antinuclei in
Fig.~\ref{tu1} by Eq.(5), are $(149\pm3)$~MeV for centrality of
0-5\%, ($142\pm 4$)~MeV for MiniBias events, and ($125 \pm 6$)~MeV
for centrality of 40-60\%, respectively. The value of the
temperature $T$ for light nuclei are close to those of antinuclei
within the error range. Obviously, since the temperature of
peripheral collision is lower then central collisions, so their
yield decrease faster with increasing of mass number $A$.

For a further analysis on the effects of centrality, we show, in
Fig.2, the yields $dN/dy$ of $d$, $\overline d$, ${_{\Lambda}^3 H}$,
$\overline{_{\overline\Lambda}^3 H}$, ${^3{He}}$
$\overline{^3{He}}$, ${^4{He}}$, and $\overline{^4{He}}$ as
functions of centrality in Au+Au collisions at
$\sqrt{s_{\rm{NN}}}=200$~GeV. All data points are normalized to the
values in the central collisions (0-5\%). The centrality dependence
of the antiproton and antilamda yield is also shown for
comparison~\cite{prl2004}. It shows that the yields of light
(anti)nuclei and (anti)hypertriton decrease with the increase of the
centrality. Furthermore, this distribution properties of light
(anti)nuclei and (anti)hypertriton production mainly depend on their
mass number, i.e. the greater the mass number is, the faster the
yield decreases.

The integrated yield is dominated by the low $p_t$ region, where
particle production is driven mainly by soft processes. The yield
per participant nucleon may reflect the formation probability of a
hadron from the bulk. We would then expect it to be sensitive to the
density of the light nuclei's constituent hadrons in the system. We
define a relative yield $R_{CY}(N_{\rm part})$ as a measure of the
(anti)nuclei production's dependence on the collision system's size
and density:
\begin{equation}
R_{\rm \small{CY}}(N_{\rm part})= \frac{({dN}/{dy})/{N_{\rm
part}}}{[({dN}/{dy})/{N_{\rm part}}]^{\rm Peripheral}}.
\end{equation}
\begin{figure}[htbp]\begin{center}
\includegraphics[width=0.65\textwidth]{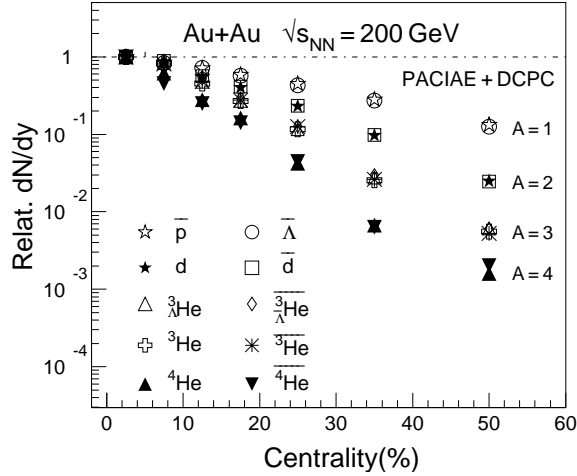}
\caption{The integrated yield $dN/dy$ of $\overline p$, $ \overline
\Lambda$, $d, \overline d$, $_{\Lambda}^3H,
\overline{_{\overline\Lambda}^3 H}$, $^3{He}, ^3{\overline{He}}$,
$^4{He}$, $^4{\overline{He}}$ in the midrapidity Au+Au collisions at
$\sqrt{s_{\rm{NN}}}=200$~GeV, normalized to the central collisions
(0-5\%), as a function of centrality.} \label{tu2}
\end{center}\end{figure}

\begin{figure*}[htbp]
\begin{center}
\includegraphics[width=0.95\textwidth]{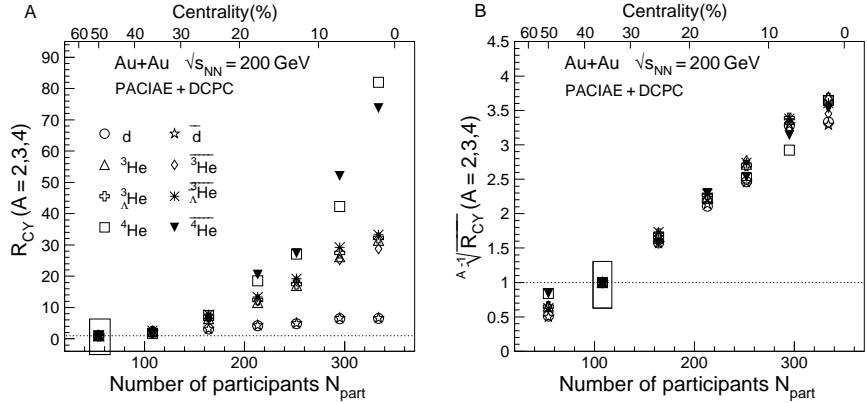}
\caption{(A) The integrated yield $dN/dy$ at midrapidity for $d,
\overline d$, $_{\Lambda}^3H, \overline{_{\overline\Lambda}^3 H}$,
$^3{He}, ^3{\overline{He}}$, $^4{He}$, $^4{\overline{He}}$ divided
by $N_{\rm part}$, normalized to the peripheral collisions
(40-60\%), plotted as a function of $N_{\rm part}$. (B) Relative
yield invariants $\sqrt[A-1]{R_{CY}}$ as a function of $N_{\rm
part}$ (cf. text for details). The results are calculated by
{\footnotesize PACIAE+DCPC} model in Au+Au collisions at
$\sqrt{s_{\rm{NN}}}=200$~GeV.} \label{tu3}
\end{center}
\end{figure*}

In Fig.~\ref{tu3}(A) we show the relative yield $R_{CY}(N_{\rm
part})$ of $d$, $\overline d$, ${_{\Lambda}^3 H}$,
$\overline{_{\overline\Lambda}^3 H}$, ${^3{He}}$
$\overline{^3{He}}$, ${^4{He}}$, and $\overline{^4{He}}$, as
functions of $N_{\rm part}$. All data points are normalized to the
values obtained in the peripheral collisions
(40-60\%)~\cite{prl2004}. It shows that the yields of light
(anti)nuclei and (anti)hypertriton per participant nucleon increase
rapidly with the increase of the number of participant
$N_{\rm{part}}$ as the $N_{\rm{part}}> 100$. Obviously, this
distribution properties of light (anti)nuclei and (anti)hypertriton
production in Au+Au collisions at $\sqrt{s_{\rm{NN}}}=200$~GeV
depend on their mass number, i.e. the greater the mass number, the
faster the increase is in yield. For a more quantitative comparison,
we define a relative yield invariants $R_{CYA}(N_{\rm part})$ as
\begin{equation}
R_{\rm \small{CYA}}(N_{\rm part})= \sqrt[A-1]{R_{\rm
\small{CY}}(N_{\rm part})},
\end{equation}
where $A$ is the atomic mass number. We show the results of
$R_{CYA}(N_{part})$, with data points normalized to the values in
the peripheral collisions (30-40\%), in Fig.3(B). We were surprised
to find that the datum  for all different (anti)nuclei and
(anti)hypertriton locate on the same straight line, i.e. the
distribution of the relative yield invariants $R_{CYA}(N_{part})$ is
independent on the number or type of the constituent hadrons.
Herein, we found a scaling property of (anti)nuclei and
(anti)hypenuclei production in relativistic heavy ion collisions. We
expect this scaling property would provide us a new clue to explore
the hadronic final state, as well as the (anti)nuclei and
(anti)hypenuclei production mechanism in relativistic heavy ion
collisions.

\section{Conclusion}

In this paper we use the {\footnotesize PACIAE+DCPC} model to
investigate the mass number dependence of light (anti)nuclei and
(anti)hypertriton production in Au+Au collisions at top RHIC energy.
The results show that the yields of light (anti)nuclei decrease
rapidly with the increase of atomic mass number, and the integrated
yield $dN/dy$ span almost 5 orders of magnitude with striking
exponential behavior which depends on the centrality. This
properties of the system can be described quantitatively by
temperature $T$ at hadronic freeze-out, and the model results are
consistent with STAR data. In addition, we studied the relative
yields per $N_{\rm part}$ of light (anti)nuclei, normalized to the
values obtained in the peripheral collisions. It is found that the
yields of light antinuclei and antihypertriton production per
participant nucleon increase linearly with $N_{\rm part}$ as
$N_{\rm{part}}> 100$, and the yield of heavy nuclei increases more
rapidly than that of light nuclei. We found that the light
(anti)nuclei and (anti)hypertriton production in Au+Au collisions at
$\sqrt{s_{\rm{NN}}}=200$~GeV exist an mass number scaling property.
We expect that this scaling characteristics can be tested in heavy
ion collision experiments.

\label{}

\begin{center} {ACKNOWLEDGMENT} \end{center}
Finally, we acknowledge the financial support from NSFC (11305144,
11303023) and Central Universities (GUGL
100237,120829,130249,130605) in China. The authors thank Nu Xu and
Ben-Hao Sa for helpful discussions.

\bibliographystyle{model6-num-names}
\bibliography{<your-bib-database>}

\begin{thebibliography}{99}
\bibitem{prl2005}  S. S. Adler et al.(PHENIX Collaboration),
Phys. Rev. Lett.,{\bf 94}, 122302 (2005).
\bibitem{star2007} B.I. Abelev et al.(STAR Collaboration), Phys. Lett. B {\bf 655}, 104 (2007).

\bibitem{star2} B. I. Abelev et al. (STAR Collaboration), Science {\bf 328}, 58
(2010).
\bibitem{star3}  H. Agakishiev et al.(STAR Collaboration), Nature {\bf 473},
353 (2011);
\bibitem{alice}
N. Sharma(ALICE Collaboration), Acta Physica Polonica B {\bf 5(2)}
605(2012).
\bibitem{alice1}
N. Sharma(ALICE Collaboration), J. Phys. G: Nucl. Part. Phys. {\bf
38}, 124189 (2011).
\bibitem{jili1} S. T. Butler and C. A. Pearson, Phys. Rev. {\bf 129}, 836 (1963).
\bibitem{jili2} A.Schwarzschild and C. Zupancic, Phys. Rev. {\bf 129}, 854 (1963).
\bibitem{jili3} H. H. Gutbrod et al., Phys. Rev. Lett. {\bf 37} 667 (1976).
\bibitem{grei}
R. Mattiello, H. Sorge, H. St\"{o}cker, and W. Greiner, Phys. Rev. C
{\bf 55}, 1443 (1997).
\bibitem{chen}
L. W. Chen and C. M. Ko, Phys. Rev. C {\bf 73}, 044903 (2006).
\bibitem{ma}
S. Zhang, J. H. Chen, H. Crawford, D. Keane, Y. G. Ma, and Z. B. Xu,
Phys. Lett. B {\bf 684}, 224 (2010).
\bibitem{pop}
V. Topor Pop and S. Das Gupta, Phys. Rev. C {\bf 81}, 054911 (2010).
\bibitem{peter}
A. Andronic, P. Braun-Munzinger, J. Stachel, and H. St\"{o}cker,
Phys.Lett. B {\bf 697}, 203 (2011).
\bibitem{ma1}
L. Xue, Y. G. Ma, J. H. Chen and S. Zhang, Phys. Rev. C {\bf 85},
064912 (2012).
\bibitem{stei}
J. Steinheimer, K. Gudima, A. Botvina, I. Mishustin, M. Bleicher,
and H. Stoecker,  Phys. Lett. {\bf B714}, 85(2012).
\bibitem{0909} B. I. Abelev et al. (The STAR Collaboration), arXiv:0909.0566
[nucl-ex].
\bibitem{ma51} H. Sato, K. Yazaki, et al., Phys. Lett. B
{\bf 98}, 153 (1981).
\bibitem{ma52} R. Scheibl and U. Heinz, Phys. Rev. C {\bf 59}, 1585
(1999).
\bibitem{prc2002} S. Albergo, R. Bellwied, M. Bennett,et al. Phys. Rev. C {\bf
65}, 034907 (2002).
\bibitem{yuyl} Y. L. Yan, G. Chen, X. M. Li, et al. Phys. Rev. {\bf C85}, 024907(2012).
\bibitem{sa2} B. H. Sa, D. M. Zhou, Y. L. Ya, et al. Comput. Phys. Commun. {\bf 183}, 333 (2012).
\bibitem{Cheng} G. Chen, Y. L. Yan, D. S. Li, et al. Phys. Rev. {\bf
C86}, 054910(2012).
\bibitem{Cheng2} G. Chen, H. Chen, J. Wu, et al. Phys. Rev. {\bf
C88}, 034908(2013).
\bibitem{prl83} T. A. Armstrong(E864 Collaboration),Phys. Rev. Lett.
{\bf 83}, 26 (1999).
\bibitem{sjo2} T. Sj\"ostrand, S. Mrenna, and P. Skands,
J. High Energy Phys. {\bf 05}, 026 (2006).
\bibitem{comb} B. L. Combridge, J. Kripfgang, and J. Ranft, Phys. Lett.
{\bf B70}, 234 (1977).
\bibitem{kubo}
K. Stowe, A introduction to thermodynamics and statistical
mechanics, Combridge, 2007; R. Kubo, Statistical Mechanics,
North-Holland Publishing Company, Amsterdam, 1965.
\bibitem{star5}J. Adams et al.(STAR Collaboration),
Phys. Rev. Lett. {\bf 98} , 062301 (2007)
\bibitem{prc2005} J.Cleymans, B.Kampfer, M.Kaneta,S.Wheaton, N.Xu, Phys. Rev. {\bf C 71}, 054901 (2005).
\bibitem{plb1995} P. Braun-Munzinger, J. Stachel, J.P. Wessels, N.
Xu, Phys. Lett. {\bf B 344}, 43 ( 1995).
\bibitem{prl1998nu} H. van Hecke, H. Sorge, and N. Xu, Phys. Rev. Lett. {\bf 81}, 5764 (1998).
\bibitem{prl2004} J. Adams et
al. (STAR Collaboration), Phys. Rev. Lett. {\bf 92}, 112301 (2004).

\bibitem{ijmpe} W. Greiner. Int. J. Mod. Phys. E: NUCL. PHYS. {\bf 5}, 1(1996).
\bibitem{ma3} Y. G. Ma, J. H. Chen, L. Xue, Front. Phys.,{\bf 7(6)}, 637
(2012).


\end{thebibliography}


\end{document}